# A Supersymmetric 3-3-1 model with MSSM -like Scalar Sector

Sutapa Sen and Aparna Dixit
Department of Physics
Christ Church P.G.College, Kanpur -208005,INDIA

## Abstract

We present a supersymmetric 3-3-1 model with exotic quarks and a charged lepton as an extension of the MSSM model with anomaly free three generations The scalar sector is studied with six triplet Higgses and the mass spectrum for light Higgs scalars are found to obey MSSM like predictions .The trilinear Higgs term in 3-3-1 is found to be consistent with the bilinear $\mu H_u H_d$ term of MSSM and play an important role in the tree-level mass spectrum of scalars.

## 1. Introduction

New physics beyond the Standard Model (SM) and its supersymmetric extensions (minimal extension of Standard Model or MSSM [1] and next-to-minimal supersymmetric extension of the Standard Model NMSSM [2,3] ) form interesting challenges to the LHC and linear colliders. An interesting class of models based on $SU(3) \otimes SU(3) \otimes U(1)$ gauge symmetry have been studied recently with supersymmetric extensions [4]. While several versions of 3-3-1 models exist in literature, these can be characterized by their embedding in larger gauge symmetry groups as $[SU(3)]^3$ [5], $SU(6) \otimes U(1)$ [6] and 3-4-1 gauge symmetry [7].

The supersymmetric version has been considered recently for 3-3-1 models [8] including a right-handed neutrino [9] and a non-exotic anti-lepton [10] .The exotic bosons for these models have non-zero lepton numbers. In this work we present a supersymmetric version of a $SU(3)_C \otimes SU(3)_L \otimes U(1)_X$ model which predicts additional exotic charged quarks and a charged lepton [11]. We consider a three-generation 3-3-1 model [12] without bilepton gauge bosons derived from 3-3-1-1 gauge symmetry which is a subgroup of $SU(4)_{PS} \otimes SU(4)_W$. The coupling constants of $SU(3)_L$ and $U(1)_X$, g and $g_X$ are related to electroweak mixing angle $\theta_W$ as $g_X^2/g^2 = \sin^2\theta_W / (1 - 4\sin^2\theta_W)$. The anomaly cancellation takes place within three generations of fermions which transform differently for the first two and the third generation. The SUSY version of the model has an extended scalar sector with three Higgs triplets along with three new scalars in adjoint representation to cancel chiral anomalies generated by Higgsinos. The mass spectrum for the Higgs scalars is obtained for neutral, pseudoscalar and charged cases. The lightest scalar $h^0_1$ is found to have a mass within the acceptable range of 114-128 GeV. Since the gauge bosons and scalars have B-L = 0, we have an analogous situation as in MSSM in which the $\mu H_u H_d$ term in superpotential is replaced by the trilinear Higgs term $\varepsilon\rho\eta\chi$ [8]. In Sec.2 we present a general formulation of the model along with the connection with MSSM and NMSSM models. In Sec.3 we present the supersymmetric 3-3-1 model along with the scalar potential and constraint equations..Sec.4 deals with the mass spectrums of the neutral scalar, pseudoscalar, single and double- charged scalars.Sec.5 deals with the numerical analysis of our work. In Sec.6 we discuss the .interactions of Higgs scalars with gauge bosons and fermions. Finally Sec.7 is a short discussion on results and conclusions.

## 2. The supersymmetric 3-3-1 model

The 3-3-1 model with exotic charged quarks and a charged lepton [9] can be embedded in $SO(12)$-derived $SU(4)_{PS} \otimes SU(4)_W$ group with $SU(4)_{PS} \to SU(3)_C \otimes U(1)_{B-L}$ [13].

In addition, $SU(4)_W \to SU(3)_L \otimes U(1)_{Y_1}$ such that $U(1)_{B-L} \otimes U(1)_{Y_1} \to U(1)_X$ gives

$3_C$-$3_L$-$1_X$ symmetry group with U(1)$_X$ charge defined by $X = Y_1 + \dfrac{(B-L)}{2}$.

The pattern of symmetry breaking in the model is given by [12]

$SU(3)_C \otimes U(1)_{B-L} \otimes SU(4)_W \to SU(3)_C \otimes SU(3)_L \otimes U(1)_{YI} \otimes U(1)_{B-L}$

$$SU(3)_C \otimes SU(3)_L \otimes U(1)_X \to SU(3)_C \otimes SU(2)_L \otimes U(1)_{X'} \otimes U(1)_X$$
$$M_\chi$$

$\to \quad SU(3)_C \times SU(2)_L \times U(1)_Y \to SU(3)_C \times U(1)_{em}$
$M_\chi \qquad\qquad M_\eta, M_\rho$ (1)

The SU(2)$_L$ weak isospin group is embedded in SU(3)$_L$ which decomposes as $SU(2)_L \otimes U(1)_{X'}$ [14] with $X' = -\sqrt{3}T_8 = -T_{3R} + Y_1$. Here $T_{3R}$ and $Y_1$ are $SU(4)_W$ generators.

The fundamental representation of $SU(3)_L \otimes U(1)_{YI}$,

$$\left(3_L, \frac{1}{2}\right) = \left(2_L, X', \frac{1}{2}\right) + \left(1_L, X', \frac{1}{2}\right) \text{ with } X' = \frac{1}{2}, -1 \ . \tag{2}$$

For $X' = -1$, we consider $T_{3R} = \dfrac{3}{2}, Y_1 = \dfrac{1}{2}$ which is different from $T_{3R} = \dfrac{1}{2}, Y_1 = -\dfrac{1}{2}$

- Under $SU(2)_L \otimes U(1)_{T_{3R}} \otimes U(1)_{Y_1}$ the fundamental triplet $3_L$ and singlets $1_L, 1_L'$ decompose as

$$3_L = \left(3_L, \frac{1}{2}\right) = \left(2_L, 0, \frac{1}{2}\right) \oplus \left(1_L, \frac{3}{2}, \frac{1}{2}\right)$$

$$1_L = \left(1_L, -\frac{1}{2}\right) = \left(1_L, -\frac{1}{2}, -\frac{1}{2}\right); 1_L' = \left(1_L, -\frac{3}{2}\right) = \left(1_L, -\frac{3}{2}, -\frac{3}{2}\right) \tag{3}$$

The electric charge operator

$$\frac{Q}{e} = T_{3L} + \sqrt{3}T_8 + \sqrt{6}T_{15} + \frac{(B-L)}{2}I_4 = T_{3L} - X' + X = T_{3L} + \frac{Y}{2} \tag{4}$$

where $T_i$, ($i = 3,8,15$) are diagonal generators of $SU(4)_W$, $Y_1 = \sqrt{6}T_{15}$, $I_4$ is the 4 x 4 identity matrix, and Y denotes hypercharge.

- **Matter multiplets:**

$$\text{Lepton } \psi_{\alpha L} = \begin{pmatrix} \nu_{l\alpha} \\ e_{l\alpha} \\ P_\alpha^+ \end{pmatrix} \sim (1_C, 3_L, 0), \quad l_\alpha = e, \mu, \tau; P_\alpha = P_e, P_\mu, P_\tau$$

$$\text{Quark } Q_i = \begin{pmatrix} d_i \\ u_i \\ D_i \end{pmatrix} \sim \left(3_C, 3^*_L, -\frac{1}{3}\right), i = 1,2 \; ; Q_3 = \begin{pmatrix} t \\ b \\ T \end{pmatrix} \sim \left(3_C, 3_L, \frac{2}{3}\right)$$

The singlet leptons are

$$l_R^C \sim (1_C, 1_L, -1); P_R^C \sim (1_C, 1_L, 1)); l = e, \mu, \tau; P = P_e, P_\mu, P_\tau.$$

The right-handed neutrinos are singlets of the model and do not contribute to anomaly cancellation. The singlet quarks

$$u_{Ri}^C \sim \left(3_C^*, 1_L, -\frac{2}{3}\right); d_{Ri}^C \sim \left(3_C^*, 1_L, \frac{1}{3}\right); D_{Ri}^C \sim \left(3_C^*, 1_L, \frac{4}{3}\right);$$

$$t_R^C \sim \left(3_C^*, 1_L, -\frac{2}{3}\right); b_R^C \sim \left(3_C^*, 1_L, \frac{1}{3}\right); T_R^C \sim \left(3_C^*, 1_L, -\frac{5}{3}\right). \tag{5}$$

- **Higgs triplets**

$$\eta = \begin{pmatrix} \eta^0 \\ \eta_1^- \\ \eta_2^+ \end{pmatrix} \sim (1,3,0); \begin{pmatrix} \eta^0 \\ \eta_1^- \end{pmatrix} \sim \left(1_C, 2_L, -\frac{1}{2}, 0\right); \eta_2^+ \sim (1_C, 1_L, 1, 0)$$

$$\rho = \begin{pmatrix} \rho^+ \\ \rho^0 \\ \rho^{++} \end{pmatrix} \sim (1,3,1); \begin{pmatrix} \rho^+ \\ \rho^0 \end{pmatrix} \sim \left(1_C, 2_L, \frac{1}{2}, 1\right); \rho^{++} \sim (1_C, 1_L, 2, 1) \qquad (6)$$

$$\chi = \begin{pmatrix} \chi^- \\ \chi^{--} \\ \chi 0 \end{pmatrix} \sim (1,3,-1); \begin{pmatrix} \chi^- \\ \chi^{--} \end{pmatrix} \sim \left(1_C, 2_L, -\frac{3}{2}, -1\right); \chi^0 \sim (1_C, 1_L, 0, -1)$$

The vacuum expectation values which are nonzero include

$$<\eta^0> = v;\ <\rho^0> = u;\ <\chi^0> = V \qquad (7)$$

We now introduce chiral superfields $\hat{\varphi}$ by extending the particle content to include squarks, sleptons and higgsinos. The superpartner for a given particle $f$ is $\tilde{f}$.

The scalar sector contains additional fermion partners or higgsinos $\tilde{\eta}$, $\tilde{\rho}$, $\tilde{\chi}$. For anomaly cancellation for higgsinos, we introduce a set of three additional scalars $\rho', \eta', \chi'$ with higgsinos $\tilde{\eta}', \tilde{\rho}', \tilde{\chi}'$ transforming in adjoint representations

$$\tilde{\eta}' = \begin{pmatrix} \tilde{\eta}'^0 \\ \tilde{\eta}_1'^+ \\ \tilde{\eta}_2'^- \end{pmatrix} \sim (1,3^*,0); \begin{pmatrix} \tilde{\eta}'^0 \\ \tilde{\eta}_1'^+ \end{pmatrix} \sim \left(1_C, 2^*_L, +\frac{1}{2}, 0\right); \tilde{\eta}_2'^- \sim (1_C, 1_L, 1, 0)$$

$$\tilde{\rho}' = \begin{pmatrix} \tilde{\rho}'^- \\ \tilde{\rho}'^0 \\ \tilde{\rho}'^{--} \end{pmatrix} \sim (1,3^*,-1); \begin{pmatrix} \tilde{\rho}'^- \\ \tilde{\rho}'^0 \end{pmatrix} \sim \left(1_C, 2^*_L, -\frac{1}{2}, -1\right); \tilde{\rho}'^{--} \sim (1_C, 1_L, -2, -1) \qquad (8)$$

$$\tilde{\chi}' = \begin{pmatrix} \tilde{\chi}'^+ \\ \tilde{\chi}'^{++} \\ \tilde{\chi}'^0 \end{pmatrix} \sim (1,3^*,+1); \begin{pmatrix} \tilde{\chi}'^+ \\ \tilde{\chi}'^{++} \end{pmatrix} \sim \left(1_C, 2^*_L, +\frac{3}{2}, +1\right); \tilde{\chi}'^0 \sim (1_C, 1_L, 0, +1)$$

The vacuum expectation values which are nonzero include

$$<\eta'> = v',\ <\rho'> = u',\ <\chi'> = V'. \qquad (9)$$

- The extended electroweak sector can be compared with minimal supersymmetric standard model (MSSM). The MSSM scalar doublet $H_d$ transforms as $(\eta^0, \eta_1^-)$ but $H_u$ is different from $(\rho^+, \rho^0)$ while $\chi^0$ is $SU(2)_L$ singlet field. These transform under $SU(2)_L \otimes U(1)_{T3R} \otimes U(1)_{Y_1}$ symmetry

$$\begin{pmatrix} H_d^0 \\ H_d^- \end{pmatrix} \sim \left(2_L, -\frac{1}{2}, 0\right); \begin{pmatrix} H_u^+ \\ H_u^0 \end{pmatrix} \sim \left(2_L, \frac{1}{2}, 0\right) \qquad (10)$$

The μ term $\varepsilon_{ab} H_d^a H_u^b \to \varepsilon_{ab} \lambda \hat{N} H_u^a H_d^b \to \varepsilon_{ijk} \kappa_1 \rho^i \eta^j \chi^k$ in the MSSM, NMSSM and present model. This is also evident in the case of pseudoscalar masses for which $\kappa_1 V$ plays the role of $m_{12}^2$ in MSSM model as will be shown later.

The superpotential and constraint equations along with superfields have been discussed in literature [8] There are 27 chiral superfields. The vector superfields for the gauge bosons of each of $SU(3)_C$, $SU(3)_L$ and $U(1)_X$ are denoted by Gluon $g^b$, gluino $\lambda^b_C$; $SU(3)_L$ gauge boson $V^b_L$ and gauginos $\lambda^b_L$ and $U(1)$ gauge boson $V_X$ and gaugino $\lambda_X$.

### 3. Higgs sector and SUSY 3-3-1

The total supersymmetric 3-3-1 Lagrangian density may be written as,

$$\mathcal{L}_{SUSY3-3-1} = \mathcal{L}_{SUSY} + \mathcal{L}_{S0FT} \qquad (11)$$

where SUSY Lagrangian density is

$$\mathcal{L}_{SUSY} = \mathcal{L}_{gauge} + \mathcal{L}_{matter} + \mathcal{L}_{scalar} \qquad (12)$$

We consider the scalar part for the present purposes

$$\mathcal{L}_{scalar} = \int d^4\theta \left[ \hat{\eta}^\dagger e^{(2g\hat{V})} \hat{\eta} + \hat{\rho}^\dagger e^{(2g\hat{V} + gx\hat{V}x)} \hat{\rho} + \hat{\chi}^\dagger e^{(2g\hat{V} - gx\hat{V}x)} \hat{\chi} \right.$$
$$\left. + \hat{\eta}'^\dagger e^{(2g\hat{V})} \hat{\eta}' + \hat{\rho}'^\dagger e^{(2g\hat{V} - gx\hat{V}x)} \hat{\rho}' + \hat{\chi}'^\dagger e^{(2g\hat{V} + gx\hat{V}x)} \hat{\chi}' \right.$$

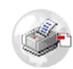

$$+ \int d^2\theta \, W + \int d^2\bar{\theta} \, \bar{W}$$

where g and $g_X$ are the gauge couplings of SU(3)$_L$ and U(1)$_X$. (13)

The superpotential $\quad W = \dfrac{W_2}{2} + \dfrac{W_3}{3}$ (14)

The bilinear terms exclude lepton number violating terms

$$W_2 = \mu_\eta \hat{\eta}\hat{\eta}' + \mu_\rho \hat{\rho}\hat{\rho}' + \mu_\chi \hat{\chi}\hat{\chi}'$$ (15)

The trilinear terms include only the following B-L conserving terms

$$W_3 \text{ (trilinear)} = \Sigma_a k^e_a \hat{L}_a \hat{\rho}' l_R{}^c + \Sigma_a k^P_a \hat{L}_a \hat{\chi}' \hat{P}_R{}^c + \Sigma_{i\alpha}[k^d_{i\alpha} \hat{Q}_i \hat{\eta} \hat{d}_{R\alpha}{}^c$$
$$+ k^u{}_{i\alpha} \hat{Q}_i \hat{\rho} \hat{u}_{R\alpha}{}^c] + \hat{Q}_3 \Sigma_\alpha [k^b{}_\alpha \hat{\rho}' \hat{d}_{R\alpha}{}^c + k^t{}_\alpha \hat{\eta}' \hat{u}_{R\alpha}{}^c]$$
$$+ k^T \hat{Q}_3 \hat{\chi}' \hat{T}_R{}^c + \Sigma_{i\beta}(k^D{}_{i\beta} \hat{Q}_i \hat{\chi} \hat{D}_{R\beta}{}^c) + f_1 \varepsilon \hat{\rho}\hat{\eta}\hat{\chi} + f_1' \varepsilon \hat{\rho}'\hat{\eta}'\hat{\chi}'$$ (16)

The indices a = 1,2,3 ; i = 1,2; $\alpha$ = 1,2,3 and $\beta$ = 1,2 refer to three generations of leptons and quarks. We consider only the scalar sector of eqn.(16). The coefficients $\mu_\rho, \mu_\eta, \mu_\chi$ have mass dimensions while all the coefficients in $W_3$ are dimensionless.

### 3.1 The Scalar Potential

The scalar potential of the model (involving $\phi_a$ = $\eta, \eta', \rho, \rho', \chi, \chi'$ fields) is given by

$$V_H = V_F + V_D + V_{soft}$$

where $V_F = -\mathcal{L}_F = \Sigma_a F_a{}^* F_a; \quad F_a{}^* = -\dfrac{\partial W}{\partial \phi_a}$ denote auxiliary fields. (17)

$$V_F = \Sigma_{ijk} [\,|\dfrac{\mu_\eta}{2}\eta_i' + \dfrac{f_1}{3}\varepsilon_{ijk}\rho_j\chi_k|^2 + |\dfrac{\mu_\rho}{2}\rho_i' + \dfrac{f_1}{3}\varepsilon_{ijk}\chi_j\eta_k|^2 + |\dfrac{\mu_\chi}{2}\chi_i' + \dfrac{f_1}{3}\varepsilon_{ijk}\rho_j\eta_k|^2$$
$$`|\dfrac{\mu_\eta}{2}\eta_i + \dfrac{f_1'}{3}\varepsilon_{ijk}\rho_j'\chi_k'|^2 + |\dfrac{\mu_\rho}{2}\rho_i + \dfrac{f_1'}{3}\varepsilon_{ijk}\chi_j\eta_k|^2 + |\dfrac{\mu_\chi}{2}\chi_i + \dfrac{f_1'}{3}\varepsilon_{ijk}\rho_j'\eta_k'|^2 \quad (18)$$

$$V_D = -\mathcal{L}_D = \frac{1}{2}(D^\alpha D^\alpha + DD), \text{ where } D^\alpha = g/2\, \varphi_i^\dagger \lambda_{ij}^\alpha \varphi_j;\quad D = g_X \varphi_i^\dagger X \varphi_i \tag{19}$$

here $\varphi = \eta, \rho, \chi$; $\lambda^\alpha (\alpha = 1,2,...8)$ are $SU(3)_L$ generators and X denotes $U(1)_X$ charge.

$$V_D = -L_D = \frac{g_X^2}{2}\left(\rho^\dagger \rho - \chi^\dagger \chi - \rho'^\dagger \rho' + \chi'^\dagger \chi'\right)^2$$
$$+ \frac{g^2}{8}\sum_{i,j}\left(\eta_i^\dagger \lambda_{ij}^a \eta_j + \rho_i^\dagger \lambda_{ij}^a \rho_j + \chi_i^\dagger \lambda_{ij}^a \chi_j - \eta_i'^\dagger \lambda_{ij}^{*a} \eta_j' - \rho_i'^\dagger \lambda_{ij}^{*a} \rho_j' - \chi_i'^\dagger \lambda_{ij}^{*a} \chi_j'\right)^2 \tag{20}$$

$$V_{soft} = -L_{soft}^{scalar} = m_\eta^2 \eta^\dagger \eta + m_\rho^2 \rho^\dagger \rho + m_\chi^2 \chi^\dagger \chi + m_{\eta'}^2 \eta'^\dagger \eta' + m_{\rho'}^2 \rho'^\dagger \rho' + m_{\chi'}^2 \chi'^\dagger \chi'$$
$$+ \kappa_1 \varepsilon \rho \eta \chi + \kappa_1' \varepsilon \rho' \eta' \chi' + H.c \tag{21}$$

where $\kappa_1, \kappa_1'$ have dimension of mass. The complete scalar potential $V_H$ given by eqn. (17) includes the neutral Higgs field $\left(X_i^0 = \rho, \eta, \chi, \rho', \eta', \chi'\right)$ terms as well as charged Higgs terms

$$V_H = \frac{g_X^2}{2}(\rho^\dagger \rho - \chi^\dagger \chi - \rho'^\dagger \rho' + \chi'^\dagger \chi')^2 + \frac{g^2}{8}\left[\frac{4}{3}\left\{(\eta^\dagger \eta)^2 + (\rho^\dagger \rho)^2 + (\chi^\dagger \chi)^2\right\} + \right.$$
$$\left.(\eta'^\dagger \eta')^2 + (\rho'^\dagger \rho')^2 + (\chi'^\dagger \chi')^2\right\} -$$

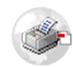

$$\frac{4}{3}\{(\eta^\dagger \eta)(\rho^\dagger\rho + \chi^\dagger\chi + \eta'^\dagger\eta' - \rho'^\dagger\rho' - \chi'^\dagger\chi') + (\rho^\dagger\rho)(\chi^\dagger\chi - \eta'^\dagger\eta' + \rho'^\dagger\rho' - \chi'^\dagger\chi')$$
$$+ (\chi^\dagger\chi)(\chi'^\dagger\chi' - \eta'^\dagger\eta' - \rho'^\dagger\rho') + (\eta'^\dagger\eta')(\rho'^\dagger\rho' + \chi'^\dagger\chi') + (\rho'^\dagger\rho')(\chi'^\dagger\chi')\}]$$
$$+ \frac{g^2}{4}\{(\eta^\dagger\rho)(\rho^\dagger\eta) + (\eta^\dagger\chi)(\chi^\dagger\eta) + (\chi^\dagger\rho)(\rho^\dagger\chi) + (\eta'^\dagger\rho')(\rho'^\dagger\eta') + (\eta'^\dagger\chi')(\chi'^\dagger\eta') + (\chi'^\dagger\rho')(\rho'^\dagger\chi')\}$$
$$+ \frac{g^2}{4}\{(\rho^\dagger\eta')(\eta^\dagger\rho') + (\chi^\dagger\eta')(\eta^\dagger\chi') + (\chi^\dagger\rho')(\rho^\dagger\chi')\}$$
$$+ (m_\eta^2 + \frac{\mu_\eta^2}{4})\eta^\dagger\eta + (m_\rho^2 + \frac{\mu_\rho^2}{4})\rho^\dagger\rho + (m_\chi^2 + \frac{\mu_\chi^2}{4})\chi^\dagger\chi +$$
$$(m_{\eta'}^2 + \frac{\mu_\eta^2}{4})\eta'^\dagger\eta' + (m_{\rho'}^2 + \frac{\mu_\rho^2}{4})\rho'^\dagger\rho' + (m_{\chi'}^2 + \frac{\mu_\chi^2}{4})\chi'^\dagger\chi' +$$
$$\frac{1}{6}\{\mu_\eta(f_1\eta'^\dagger\rho\chi + f_1'\eta^\dagger\rho'\chi') + \mu_\rho(f_1\rho'^\dagger\chi\eta + f_1'\rho^\dagger\chi'\eta') - \mu_\chi(f_1\chi'^\dagger\rho\eta + f_1'\chi^\dagger\rho'\eta')\} +$$
$$\frac{|f_1|^2}{9}\{(\rho^\dagger\rho)(\chi^\dagger\chi) + (\chi^\dagger\chi)(\eta^\dagger\eta) + (\rho^\dagger\rho)(\eta^\dagger\eta)\} + \frac{|f_1'|^2}{9}\{(\rho'^\dagger\rho')(\chi'^\dagger\chi') + (\chi'^\dagger\chi')(\eta'^\dagger\eta') +$$
$$(\rho'^\dagger\rho')(\eta'^\dagger\eta')\} + (\kappa_1\rho\eta\chi + \kappa_1'\rho'\eta'\chi') + H.c.$$

(22)

We introduce the expansion of neutral scalar fields

$X_i^0 = \frac{1}{\sqrt{2}}(v_{Xi} + \xi_{Xi} + i\zeta_{Xi})$ where the vacuum expectation values [VEV] include

$v = v_\eta = \langle\eta^0\rangle$, $u = v_\rho = \langle\rho^0\rangle$, $V = v_\chi = \langle\chi^0\rangle$ ;

$v' = v_{\eta'} = \langle\eta'^0\rangle, u' = v_{\rho'} = \langle\rho'^0\rangle, V' = v_{\chi'} = \langle\chi'^0\rangle$  (23)

The parameter tan β = u /v corresponds to the MSSM parameter tanβ = <h$_2$>/<h$_1$> of vev's of two Higgs doublets. .The CP even real fields are $\xi_{Xi}$ while CP-odd imaginary fields are $\zeta_{Xi}$.

### 3.2 The constraint equations in 3-3-1 model

The requirement of vanishing of linear terms in fields, $\dfrac{\partial V_H^{min}}{\partial v_{Xi}} = 0$ give the constraint equations

$$m_\eta^2 + \frac{\mu_\eta^2}{4} = -\frac{g^2}{12}\left[2v^2 - v'^2 - u^2 + u'^2 - V^2 + V'^2\right] -$$

$$\frac{1}{6v\sqrt{2}}\left[\mu_\rho f_1 u'V - \mu_\chi f_1 uV' + \mu_\eta f_1' u'V'\right]$$

$$-\frac{|f_1|^2}{18}(V^2 + u^2) - \frac{\kappa_1}{2\sqrt{2}}\frac{uV}{v} \qquad (24)$$

$$m_\rho^2 + \frac{\mu_\rho^2}{4} = -\frac{g^2}{12}\left[2u^2 - u'^2 - v^2 + v'^2 - V^2 + V'^2\right] - \frac{g_X^2}{2}\left[u^2 - u'^2 - V^2 + V'^2\right]$$

$$-\frac{1}{6u\sqrt{2}}\left[\mu_\eta f_1 v'V - \mu_\chi f_1 vV + \mu_\rho f_1' v'V'\right] - \frac{|f_1|^2}{18}(V^2 + v^2) - \frac{\kappa_1}{2\sqrt{2}}\frac{vV}{u} \qquad (25)$$

$$m_\chi^2 + \frac{\mu_\chi^2}{4} = -\frac{g^2}{12}\left[2V^2 - V'^2 - u^2 + u'^2 - v^2 + v'^2\right] - \frac{g_X^2}{2}\left[V^2 - V'^2 - u^2 + u'^2\right]$$

$$-\frac{1}{6\sqrt{2}V}\left[\mu_\eta f_1 v'u - \mu_\chi f_1' u'v' + \mu_\rho f_1 u'v\right] - \frac{|f_1|^2}{18}(u^2 + v^2) - \frac{\kappa_1}{2\sqrt{2}}\frac{uv}{V} \qquad (26)$$

$$m_{\eta'}^2 + \frac{\mu_\eta^2}{4} = -\frac{g^2}{12}\left[2v'^2 - v^2 + u^2 - u'^2 + V^2 - V'^2\right] -$$

$$\frac{1}{6v'\sqrt{2}}\left[\mu_\rho f_1' uV' - \mu_\chi f_1 u'V + \mu_\eta f_1 uV\right]$$

$$-\frac{|f_1'|^2}{18}(V'^2 + u'^2) - \frac{\kappa_1'}{2\sqrt{2}}\frac{u'V'}{v'} \qquad (27)$$

$$m_{\rho'}^2 + \frac{\mu_\rho^2}{4} = -\frac{g^2}{12}\left[2u'^2 - u^2 - v'^2 + v^2 + V^2 - V'^2\right] - \frac{g_X^2}{2}\left[u'^2 - u^2 + V^2 - V'^2\right]$$

$$-\frac{1}{6u'\sqrt{2}}\left[\mu_\eta f_1 vV' - \mu_\chi f_1' v'V + \mu_\rho f_1 vV\right] - \frac{|f_1'|^2}{18}(V'^2 + v'^2) - \frac{\kappa_1'}{2\sqrt{2}}\frac{v'V'}{u'} \qquad (28)$$

$$m_{\chi'}^2 + \frac{\mu_\chi^2}{4} = -\frac{g^2}{12}\left[2V'^2 - V^2 + u^2 - u'^2 + v^2 - v'^2\right] - \frac{g_X^2}{2}\left[V'^2 - V^2 + u^2 - u'^2\right]$$

$$-\frac{1}{6\sqrt{2}V'}\left[\mu_\eta f_1' vu - \mu_\chi f_1 uv + \mu_\rho f_1' uv'\right] - \frac{|f_1'|^2}{18}(u'^2 + v'^2) - \frac{\kappa_1'}{2\sqrt{2}}\frac{u'v'}{V'} \qquad (29)$$

To obtain mass terms, we use terms quadratic in scalars

$$M_{ij}^2 = \frac{\partial^2 V_H^{min}}{\partial \phi_i \partial \phi_j} \qquad (30)$$

where $\phi_i = \eta, \eta', \rho, \rho', \chi, \chi'$.

We introduce the following parameters which show the deviation of the model from MSSM predictions for scalar masses.

$$X_\eta = \frac{1}{6v\sqrt{2}} \left[ \mu_\rho f_1 u'V - \mu_\chi f_1 uV' + \mu_\eta f_1' u'V' \right];$$

$$X'_\eta = \frac{1}{6v'\sqrt{2}} \left[ \mu_\rho f_1' uV' - \mu_\chi f_1' u'V + \mu_\eta f_1 uV \right]$$

$$X_\rho = \frac{1}{6u\sqrt{2}} \left[ \mu_\eta f_1 v'V - \mu_\chi f_1 vV + \mu_\rho f_1' v'V' \right];$$

$$X'_\rho = \frac{1}{6u'\sqrt{2}} \left[ \mu_\eta f_1' vV' - \mu_\chi f_1' v'V' + \mu_\rho f_1 vV \right]$$

$$X_\chi = \frac{1}{6\sqrt{2}V} \left[ \mu_\eta f_1 v'u - \mu_\chi f_1' u'v' + \mu_\rho f_1 u'v \right];$$

$$X'_\chi = \frac{1}{6\sqrt{2}V'} \left[ \mu_\eta f_1' vu' - \mu_\chi f_1 uv + \mu_\rho f_1' uv' \right] \qquad (31)$$

## 4. Mass spectrum in the neutral scalar sector

The neutral scalar CP-even sector do not have Goldstone bosons and include three pairs of massive physical fields. In the basis of $(\xi_\eta, \xi_\rho, \xi_{\eta'}, \xi_{\rho'}, \xi_\chi, \xi_{\chi'})$ after imposing the constraint equations we obtain 6 x 6 mass matrix

$$M_H^2 = \begin{pmatrix} M^2_{4\eta\rho} & 0 \\ 0 & M^2_{2\chi\chi'} \end{pmatrix} \quad \text{where the 4 x 4 and 2 x 2 submatrices are}$$

$$M^2_{4\eta\rho} = \begin{pmatrix} M_{\eta\eta}^2 & \cdots & M_{\eta\rho'}^2 \\ \vdots & \ddots & \vdots \\ M_{\rho'\eta}^2 & \cdots & M_{\rho\rho'}^2 \end{pmatrix} \quad \text{and} \quad M^2_{2\chi\chi'} = \begin{pmatrix} M^2_{\chi\chi} & M^2_{\chi\chi'} \\ M^2_{\chi'\chi} & M^2_{\chi'\chi'} \end{pmatrix}$$

We consider the 4x4 mass matrix $M^2_{4\eta\rho}$ as two submatrices in the bases of ($\xi_\eta, \xi_\rho$) and ($\xi_{\eta'}, \xi_{\rho'}$) which give two pairs of massive neutral CP-even physical Higgs fields ($H^0_1$, $h^0_1$) and ($H^0_2$, $h^0_2$).

In terms of the mixing angles $\alpha_1, \alpha_2$

$$\begin{aligned} H^0_1 &= \cos\alpha_1 \xi_\eta + \sin\alpha_1 \xi_\rho & H^0_2 &= \cos\alpha_2 \xi_{\eta'} + \sin\alpha_2 \xi_{\rho'} \\ h^0_1 &= -\sin\alpha_1 \xi_\eta + \cos\alpha_1 \xi_\rho & h^0_2 &= -\sin\alpha_2 \xi_{\eta'} + \cos\alpha_2 \xi_{\rho'} \end{aligned} \quad (32)$$

The matrix elements of $M^2_{4\eta\rho}$ include

$$M^2_{\eta\eta} = \frac{g^2}{3}v^2 - X_\eta - \frac{\kappa_1}{2\sqrt{2}}\frac{uV}{v} = A$$

$$M^2_{\eta\rho} = -\frac{g^2}{6}uv + \frac{|f_1|^2}{9}uv - \frac{\mu_\chi f_1 V'}{6\sqrt{2}} + \frac{\kappa_1}{2\sqrt{2}}V = B$$

$$M^2_{\rho\rho} = \left(\frac{g^2}{3} + g_X^2\right)u^2 - X_\rho - \frac{\kappa_1}{2\sqrt{2}}\frac{vV}{u} = C \qquad (33)$$

The physical eigenvalues

$$m^2_{h1} = \frac{1}{2}\left(M_1 - \sqrt{M_1^2 + M_1'}\right); m^2_{H_1} = \frac{1}{2}\left(M_1 + \sqrt{M_1^2 + M_1'}\right) \qquad (34)$$

$$M_1 = A + C; M_1' = -4AC + B^2$$

$$\tan 2\alpha_1 = \frac{2B}{A - C}$$

The second pair of Higgses ($H_2^0, h_2^0$) are obtained from matrix elements

$$M^2_{\eta'\eta'} = \frac{g^2}{3}v'^2 - X'_\eta - \frac{\kappa'_1}{2\sqrt{2}}u'\frac{V'}{v'} = A'$$

$$M^2_{\eta'\rho'} = -\frac{g^2}{6}u'v' + \frac{|f'_1|^2}{9}u'v' - \frac{\mu_\chi f'_1 V}{6\sqrt{2}} + \frac{\kappa'_1}{2\sqrt{2}}V' = B'$$

$$M^2_{\rho'\rho'} = \left(\frac{g^2}{3} + g_X^2\right)u'^2 - X'_\rho - \frac{\kappa'_1}{2\sqrt{2}}\frac{v'V'}{u'} = C' \tag{35}$$

The physical eigenvalues

$$m^2_{h2} = \frac{1}{2}\left(M_2 - \sqrt{M_2^2 + M'_2}\right); m^2_{H2} = \frac{1}{2}\left(M_2 + \sqrt{M_2^2 + M'_2}\right)$$
$$M_2 = A' + C'; M'_2 = -4A'C' + B'^2$$

$$\tan 2\alpha_2 = \frac{2B'}{A' - C'} \tag{36}$$

For the third pair of neutral scalars ($H_3^0, h_3^0$) we consider the matrix elements of $M^2_{2\chi\chi'}$,

$$M^2_{\chi\chi} = \left(\frac{g^2}{3} + g_X^2\right)V^2 - X_\chi - \frac{\kappa_1}{2\sqrt{2}}\frac{uv}{V} = A''$$

$$M^2_{\chi\chi'} = \left(\frac{g^2}{6} + g_X^2\right)VV' = B'' \tag{37}$$

$$M^2_{\chi'\chi'} = \left(\frac{g^2}{3} + g_X^2\right)V'^2 - X'_\chi - \frac{\kappa'_1}{2\sqrt{2}}\frac{u'v'}{V'} = C''$$

The physical fields are

$$H_3^0 = \cos\alpha_3\ \xi_\chi + \sin\alpha_3\ \xi'_\chi\ ;\ h_3^0 = -\sin\alpha_3\ \xi_\chi + \cos\alpha_3\ \xi'_\chi \tag{38}$$

The physical eigenvalues

$$m^2_{h3} = \frac{1}{2}\left(M_3 - \sqrt{M_3^2 + M'_3}\right); m^2_{H3} = \frac{1}{2}\left(M_3 + \sqrt{M_3^2 + M'_3}\right)$$
$$M_3 = A'' + C''; M'_3 = -4A''C'' + B''^2$$

$$\tan 2\alpha_3 = \frac{2B''}{A'' - C''} \tag{39}$$

An interesting correspondence with MSSM soft bilinear term is that $\frac{\kappa_1}{2\sqrt{2}}V$ now plays the role of $m_{12}^2$ in the matrix elements for masses of lightest scalar Higgses.

## 4.1 Mass spectrum in neutral pseudoscalar sector

The 6 x 6 mass square matrix in the pseudoscalar sector in bases $(\zeta_\eta, \zeta_\rho, \zeta_{\eta'}, \zeta_{\rho'}, \zeta_\chi, \zeta_{\chi'})$

$$M_{PH}^2 = \begin{pmatrix} M^2_{4\eta\rho} & 0 \\ 0 & M^2_{2\chi\chi'} \end{pmatrix} \quad \text{where the 4 x 4 and 2 x 2 submatrices are}$$

$$M^2_{4\eta\rho} = \begin{pmatrix} M^2_{\eta\eta} & M^2_{\eta\rho} & M^2_{\eta\eta'} & M^2_{\eta\rho'} \\ M^2_{\rho\eta} & M^2_{\rho\rho} & M^2_{\rho\eta'} & M^2_{\rho\rho'} \\ M^2_{\eta'\eta} & M^2_{\eta'\rho} & M^2_{\eta'\eta'} & M^2_{\eta'\rho'} \\ M^2_{\rho'\eta} & M^2_{\rho'\rho} & M^2_{\rho'\eta'} & M^2_{\rho'\rho'} \end{pmatrix} \quad \text{and} \quad M^2_{2\chi\chi'} = \begin{pmatrix} M^2_{\chi\chi} & M^2_{\chi\chi'} \\ M^2_{\chi'\chi} & M^2_{\chi'\chi'} \end{pmatrix} \quad (40)$$

We consider the 4 x 4 matrix $M_{4\eta\rho}^2$ as two 2 x 2 submatrices with bases $(\zeta_\eta, \zeta_\rho), (\zeta_{\eta'}, \zeta_{\rho'})$. In the first basis, $(\zeta_\eta, \zeta_\rho)$

$$M^2_{\zeta_\eta\zeta_\rho} = \begin{pmatrix} A_P & B_P \\ B_P & C_P \end{pmatrix}$$

$$M^2_{\eta\eta} = -X_\eta - \frac{\kappa_1}{2\sqrt{2}} \frac{uV}{v} = A_P$$

$$M^2_{\eta\rho} = \frac{\mu_\chi f_1}{6\sqrt{2}} V' - \frac{\kappa_1}{2\sqrt{2}} V = M^2_{\rho\eta} = B_P \quad (41)$$

$$M^2_{\rho\rho} = -X_\rho - \frac{\kappa_1}{2\sqrt{2}} \frac{vV}{u} = C_P$$

The non-vanishing trace and a vanishing determinant imply a massless Goldstone boson and a massive CP-odd pseudoscalar . Thus physical fields include a neutral CP-odd pseudoscalar $A_1$ and a Goldstone boson $G_1^0$.

$A_1 = \sin\beta_1 \zeta_\eta + \cos\beta_1 \zeta_\rho; \ G_1^0 = -\cos\beta_1 \zeta_\eta + \sin\beta_1 \zeta_\rho$

$$\tan 2\beta_1 = \frac{2B_P}{A_P - C_P} \ ; \ m^2_{A1} = A_P + C_P \qquad (42)$$

In the second basis $(\zeta_{\eta'}, \zeta_{\rho'})$

$$m^2_{\eta'\rho'} = \begin{pmatrix} A'_P & B'_P \\ B'_P & C'_P \end{pmatrix}$$

$$M^2_{\eta'\eta'} = -X'_\eta - \frac{\kappa'_1}{2\sqrt{2}} \frac{u'V'}{v'} = A'_P$$

$$M^2_{\eta'\rho'} = \frac{\mu_\chi f'_1}{6\sqrt{2}} V - \frac{\kappa'_1}{2\sqrt{2}} V' = B'_P \qquad (43)$$

$$M^2_{\rho'\rho'} = -X'_\rho - \frac{\kappa'_1}{2\sqrt{2}} \frac{v'V'}{u'} = C'_P$$

The physical fields include a neutral CP-odd pseudoscalar $A_2$ and a Goldstone boson $G_2^0$

$A_2 = \sin\beta_2 \zeta_{\eta'} + \cos\beta_2 \zeta_{\rho'}; \ G_2^0 = -\cos\beta_2 \zeta_{\eta'} + \sin\beta_2 \zeta_{\rho'}$

$$\tan 2\beta_2 = \frac{2B'_P}{A'_P - C'_P} \ ; \ m^2_{A_2} = A'_P + C'_P \qquad (44)$$

$$M^2_{\eta\eta'} = M^2_{\rho\rho'} = 0; M^2_{\eta\rho'} = \frac{\mu_\rho f_1}{6\sqrt{2}} V + \frac{\mu_\eta f'_1}{6\sqrt{2}} V'; M^2_{\rho\eta'} = \frac{\mu_\eta f_1}{6\sqrt{2}} V + \frac{\mu_\rho f'_1}{6\sqrt{2}} V'; M^2_{\eta'\rho'} = \frac{\mu_\chi f'_1}{6\sqrt{2}} V - \frac{\kappa'_1}{2\sqrt{2}} V'$$

$$(45)$$

From the submatrix $M^2_{2\chi\chi'}$ with matrix elements

$$M^2_{\chi\chi} = -X_\chi - \frac{\kappa_1}{2\sqrt{2}}\frac{uv}{V}$$

$$M^2_{\chi\chi'} = B_P = 0 \tag{46}$$

$$M^2_{\chi'\chi'} = -X'_\chi - \frac{\kappa'_1}{2\sqrt{2}}\frac{u'v'}{V'}$$

we obtain masses for two pseudoscalars $A_3, A_3'$ directly

## 4.2 Mass spectrum for single charged scalar

In the single charged sector, two charged Goldstone bosons are obtained with one each for masses of ordinary and exotic gauge bosons along with six massive scalars.

The basis for 8 x 8 mass square matrix is $(\eta_1^+, \rho^+, \eta_1'^+, \rho'^+, \eta_2^+, \chi^+, \eta_2'^+, \chi'^+)$

$$M^2_{sc} = \begin{pmatrix} M^2_{4\eta\rho} & & \\ & M^2_{2\eta_2\chi} & \\ & & M^2_{2\eta_2'\chi'} \end{pmatrix}$$

where 4 x 4 matrix

$$M^2_{4\eta\rho} = \begin{pmatrix} M^2_{\eta_1^+\eta_2^-} & M^2_{\eta_1^+\rho^-} & M^2_{\eta_1^+\eta_1'^-} & M^2_{\eta_1^+\rho'^-} \\ & M^2_{\rho^+\rho^-} & M^2_{\rho^+\eta_1'^-} & M^2_{\rho^+\rho'^-} \\ & & M^2_{\eta_1'^+\eta_1'^-} & M^2_{\eta_1'^+\rho'^-} \\ & & & M^2_{\rho'^+\rho'^-} \end{pmatrix} \tag{47}$$

$$M^2_{\eta_1^+\eta_1^-} = \frac{g^2}{8}u^2 - X_\eta - \frac{\kappa_1}{2\sqrt{2}}\frac{uV}{v}$$

$$M^2_{\rho^+\rho^-} = \frac{g^2}{8}v^2 - X_\rho - \frac{\kappa_1}{2\sqrt{2}}\frac{vV}{u}$$

$$M^2_{\eta_1'^+\eta_1'^-} = \frac{g^2}{8}u'^2 - X'_\eta - \frac{\kappa'_1}{2\sqrt{2}}\frac{u'V'}{v'}$$

$$M^2_{\rho'^+\rho'^-} = \frac{g^2}{8}v'^2 - X'_\rho - \frac{\kappa'_1}{2\sqrt{2}}\frac{v'V'}{u'}$$

The off-diagonal matrix elements

$$M^2_{\eta_1^+\rho^-} = \frac{g^2}{8}(vu+v'u') + \frac{\mu_\chi f_1 V'}{6\sqrt{2}} - \kappa_1 \frac{V}{2\sqrt{2}}$$

$$M^2_{\eta_1^+\eta_1'^-} = 0;\ M^2_{\eta_1^+\rho'^-} = -\frac{\mu_\eta f_1'}{6\sqrt{2}}V';\ M^2_{\rho^+\eta_1'} = -\frac{\mu_\eta f_1 V}{6\sqrt{2}};\ M^2_{\rho^+\rho'^-} = 0 \qquad (48)$$

$$M^2_{\eta_1'^+\rho'^-} = \frac{g^2}{8}(vu+v'u') + \frac{\mu_\chi f_1' V}{6\sqrt{2}} - \kappa'_1 \frac{V'}{2\sqrt{2}}$$

We consider the base ($\eta_1^+, \rho^+$) for which 2 x 2 mass-squared matrix gives non-vanishing trace and vanishing determinant. The physical fields include a charged Goldstone boson and massive charged Higgs $\left(G_1^\pm, H_1^\pm\right)$ where

$$H_1^\pm = \sin\gamma_1 \eta_1^\pm + \cos\gamma_1 \rho^\pm;\ G_1^\pm = -\cos\gamma_1 \eta_1^\pm + \sin\gamma_1 \rho^\pm$$

$$m^2_{H_1^\pm} = M^2_{\eta_1^+\eta_1^-} + M^2_{\rho^+\rho^-};\ \tan 2\gamma_1 = \frac{2M^2_{\eta_{12}^+\rho^-}}{M^2_{\eta_1^+\eta_1^-} - M^2_{\rho^+\rho^-}} \qquad (49)$$

For the base ($\eta_1'^+, \rho'^+$) the 2 x 2 mass-squared matrix elements in eqn.(48)

$$M^2_{\eta_1'^+\eta_1'^-} = A^+,\ M^2_{\rho'^+\rho'^-} = C^+,\ M^2_{\eta_1'^+\rho'^-} = B^+ = M^2_{\rho'^-\eta_1'^+} \qquad (50)$$

Two massive charged scalars $H_2^\pm, h_2^\pm$ are obtained with mixing angle $\gamma_2$,

$$\tan 2\gamma_2 = \frac{2B^+}{A^+ - C^+}$$

$$\begin{aligned} H_2^\pm &= \cos\gamma_2 \eta_1'^\pm + \sin\gamma_2 \rho'^\pm \\ h_2^\pm &= -\sin\gamma_2 \eta_1'^\pm + \cos\gamma_2 \rho'^\pm \end{aligned} \qquad (51)$$

$m^2_{H_2^\pm}, m^2_{h_2^\pm}$ are obtained as before, where $H_2^\pm$ is heavier than $h_2^\pm$.

From the 2 x 2 submatrix

$$M^2_{2\eta_2\chi} = \begin{pmatrix} M^2_{\eta_2^+\eta_2^-} & M^2_{\eta_2^+\chi^-} \\ M^2_{\chi^+\eta_2^-} & M^2_{\chi^+\chi^-} \end{pmatrix} \tag{52}$$

$$M^2_{\eta_2^+\eta_2^-} = \frac{g^2}{8}V^2 - X_\eta - \frac{\kappa_1}{2\sqrt{2}}\frac{uV}{v}$$

$$M^2_{\eta_2^+\chi^-} = \frac{g^2}{8}(vV + v'V') - \frac{\mu_\rho f_1 u'}{6\sqrt{2}} - \kappa_1 \frac{u}{\sqrt{2}} \quad ; \quad M^2_{\chi^+\chi^-} = \frac{g^2}{8}v^2 - X_\chi - \frac{\kappa_1}{2\sqrt{2}}\frac{uv}{V}$$

The physical fields are Goldstone boson $G_3^+$ and massive Higgs $H_3^+$, mixing angle $\gamma_3$

$$\tan 2\gamma_3 = 2\frac{M^2_{\eta_2^+\chi^-}}{M^2_{\eta_2^+\eta_2^-} - M^2_{\chi^+\chi^-}}$$

$$H_3^\pm = \sin\gamma_3 \eta_2^\pm + \cos\gamma_3 \chi^\pm ; G_3^\pm = -\cos\gamma_3 \eta_2^\pm + \sin\gamma_3 \chi^\pm$$

The mass of $H_3^+$ is obtained as

$$m^2_{H_3^\pm} = M^2_{\eta_2^+\eta_2^-} + M^2_{\chi^+\chi^-} \tag{53}$$

The matrix $M^2_{2\eta_2'\chi'} = \begin{pmatrix} M^2_{\eta_2'^+\eta_2'^-} & M^2_{\eta_2'^+\chi'^-} \\ M^2_{\chi'^+\eta_2'^-} & M^2_{\chi'^+\chi'^-} \end{pmatrix}$

$$M^2_{\eta_2'^+\eta_2'^-} = \frac{g^2}{8}V'^2 - X'_\eta - \frac{\kappa_1'}{2\sqrt{2}}\frac{u'V'}{v'} = A_P'$$

$$M^2_{\eta_2'^+\chi'^-} = \frac{g^2}{8}(vV + v'V') - \frac{\mu_\rho f_1' u}{6\sqrt{2}} - \kappa_1' \frac{u'}{\sqrt{2}} = B_P'$$

$$M^2_{\chi'^+\chi'^-} = \frac{g^2}{8}v'^2 - X'_\chi - \frac{\kappa_1'}{2\sqrt{2}}\frac{u'v'}{V'} = C_P' \tag{54}$$

Mass eigenvalues of two massive charged scalars $H_4^\pm$, $h_4^\pm$ are obtained as before.

$$H_4^\pm = \cos\gamma_4 \eta_2'^\pm + \sin\gamma_4 \chi'^\pm ; h_4^\pm = -\sin\gamma_4 \eta_2'^\pm + \cos\gamma_4 \chi'^\pm ;$$

$$\tan 2\gamma_4 = \frac{2B'_P}{\left(A'_P - C'_P\right)}$$

## 4.3 Mass spectrum for double charged scalars

The spectrum of doubly charged scalars include one doubly charged goldstone boson and three massive physical fields. The squared mass matrix in the basis $\{\rho^{++}, (\chi^{--})^*, (\rho'^{--})^*, \chi'^{++}\}$

$$M^2_{dc} = \begin{pmatrix} M^2_{\rho^{++}\rho^{--}} & M^2_{\rho^{++}\chi^{--}} & 0 & M^2_{\rho^{++}\chi'^{--}} \\ & M^2_{\chi^{++}\chi^{--}} & M^2_{\chi^{++}\rho'^{--}} & 0 \\ & & M^2_{\rho'^{++}\rho'^{--}} & M^2_{\rho'^{++}\chi'^{--}} \\ & & & M^2_{\chi'^{++}\chi'^{--}} \end{pmatrix} \quad (55)$$

$$M^2_{\rho^{++}\rho^{--}} = \frac{g^2}{8}V^2 - X_\rho - \frac{\kappa_1}{2\sqrt{2}}\frac{vV}{u}$$

$$M^2_{\rho^{++}\chi^{--}} = \frac{g^2}{8}(uV + u'V') - \mu_\eta f_1 \frac{v'}{6\sqrt{2}} - \kappa_1 \frac{v}{\sqrt{2}} = B^{++} = M^2_{\chi^{--}\rho^{++}}$$

$$M^2_{\chi^{++}\chi^{--}} = \frac{g^2}{8}u^2 - X_\chi - \frac{\kappa_1}{2\sqrt{2}}\frac{uv}{V}$$

$$M^2_{\chi^{++}\rho'^{--}} = -\frac{\mu_\rho f_1 v}{6\sqrt{2}} \quad ; \quad M^2_{\rho^{++}\chi'^{--}} = \frac{\mu_\chi f_1 v}{6\sqrt{2}}$$

$$M^2_{\rho'^{++}\rho'^{--}} = \frac{g^2}{8}V'^2 - X'_\rho - \frac{\kappa'_1}{2\sqrt{2}}\frac{v'V'}{u'}$$

$$M^2_{\rho'^{++}\chi'^{--}} = \frac{g^2}{8}(uV+u'V') - \mu_\eta f_1' \frac{v}{6\sqrt{2}} - \kappa_1' \frac{v'}{\sqrt{2}}$$

$$M^2_{\chi'^{++}\chi'^{--}} = \frac{g^2}{8}u'^2 - X_\chi' - \frac{\kappa_1'}{2\sqrt{2}} \frac{u'v'}{V'}$$

We consider a vanishing determinant and non-vanishing trace of 2 x 2 matrix in the base ($\rho^{++}, \chi^{--*}$) The scalars iinclude Goldstone boson $G^{\pm\pm}$ and a physical charged Higgs $H^{\pm\pm}$ whose mass eigenvalue is

$$m^2_{H^{\pm\pm}} = M^2_{\rho^{++}\rho^{--}} + M^2_{\chi^{++}\chi^{--}}; \quad \tan 2\theta_1 = \frac{2M^2_{\rho^{++}\chi^{--}}}{M^2_{\rho^{++}\rho^{--}} - M^2_{\chi^{++}\chi^{--}}} \quad (56)$$

$H^{++} = \rho^{++}\sin\theta_1 + \chi^{++}\cos\theta_1$

$G^{++} = -\rho^{++}\cos\theta_1 + \chi^{++}\sin\theta_1$

From the 2 x 2 matrix in base ($\rho'^{++}, \chi'^{++}$), two massive scalars ($H'^{++}, h'^{++}$) are obtained where $H'^{++}$ is heavier than $h'^{++}$.

$H'^{++} = \rho'^{++}\sin\theta_2 + \chi'^{++}\cos\theta_2$

$H'^{++} = -\rho'^{++}\cos\theta_2 + \chi'^{++}\sin\theta_2$ (57)

The masses are obtained by diagonalization as before.

**Constraints on Higgs scalar masses:**

In analogy with MSSM, we define $\tan\beta = u/v$ with the constraint

$$v^2 + u^2 + v'^2 + u'^2 = (246\text{ GeV})^2. \quad (58)$$

This follows from the gauge boson masses [11] The relation for scalar mass-squared

$$m^2_{H^\pm} > m^2_{A_1}; \quad m^2_{h_1} + m^2_{H_1} > m^2_{A_1}$$

## 5. Numerical analysis

The numerical assignments for the various parameters used in obtaining the mass spectrum are as follows

$f_1 = 1$, $f_1^{/} = 10^{-6}$ are dimensionless parameters; $\kappa_1 = -10 GeV = -\kappa_1'$ ;

$-\mu_\eta = \mu_\rho = \mu_\chi = 1000 GeV$ ;

The constraint on vev's , $u^2 + u'^2 + v^2 + v'^2 = (246 GeV)^2$ is used along with

$\tan\beta = \dfrac{u}{v}; u' = v' = V' = 1 GeV$.

The variation of the lightest scalar masses with tanβ is plotted for V= 1000GeV in graph 1. We find the condition $m_{H_1^{\pm}} > m_{A1}$ to be satisfied at all values of tanβ along with the second condition. The value of the mass of lightest neutral Higgs is in the range of 114-128 GeV at tanβ = 4 – 5, with V=1TeV. In figs.2 to 4 we plot Higgs masses vs.V at tanβ =5. The value for Higgs masses at V ≤ 2TeV are found to be within experimental ranges. Since κ₁V plays the role of Bμ in MSSM, a value of 10-15 TeV is obtained for this parameter at tan β = 5. The model does not allow large values for u /v within these ranges.

## 6. Higgs couplings to Fermions

The Yukawa Lagrangians that respects the 3-3-1 gauge symmetry are given by

$$L_Y^q = \sum_{i=1,2} Q_{iL} \left[ \sum_{\alpha=1}^{3} \left( \kappa_{i\alpha}^d \eta \bar{d}_{R\alpha} + \kappa_{i\alpha}^u \rho \bar{u}_{R\alpha} \right) + \sum_{\beta=1,2} \kappa_{i\beta}^D \chi \bar{D}_{R\beta} \right] + Q_{3L} \left[ \sum_{\alpha=1}^{3} \left( \kappa_{\alpha}^b \rho' \bar{d}_{R\alpha} + \kappa_{\alpha}^t \eta' \bar{u}_{R\alpha} \right) + \kappa^T \chi' \bar{T}_R \right]$$

$$-L_Y^l = \sum_{a=1}^{3} L_{aL} \left[ \sum_{b=1}^{3} \left( \kappa_{ab}^e \rho' e_{Rb}^c + \kappa_{ab}^P \chi' P_{Rb}^c \right) \right]$$

Here $\kappa^q, \kappa^l$ are Yukawa coupling constants which are expressed in terms of fermion masses and vev's of Higgs fields as

$$\kappa^d_{i\alpha} = \sqrt{2}\frac{m_d}{v}; \kappa^u_{i\alpha} = \sqrt{2}\frac{m_u}{u}; \kappa^b_{i\alpha} = \sqrt{2}\frac{m_b}{u'}; \kappa^t_{i\alpha} = \sqrt{2}\frac{m_t}{v'}; \kappa^D_{i\alpha} = \sqrt{2}\frac{m_D}{V}; \kappa^T_{i\alpha} = \sqrt{2}\frac{m_T}{V'}$$

$$\kappa^e_{ab} = \sqrt{2}\frac{m_e}{u'}; \kappa^P_{ab} = \sqrt{2}\frac{m_P}{V'}$$

Unlike MSSM, the third generation quarks couple to adjoint scalars $\rho'$, $\eta'$ which give the massive physical Higgses ($h_2$, $H_2$, $A_2$) distinct from the lightest Higgs scalars ($h_1$, $H_1$, $A_1$)

The Feynman rules for Higgs couplings for third generation quarks are

$$t_L t_R^c\ h_i^0:\quad -m_t \frac{\left(\cos\alpha_2 H_2^0 - \sin\alpha_2 h_2^0\right)}{\sqrt{2}v'};\quad b_L b_R^c h_i^0: -m_b \frac{\left(\sin\alpha_2 H_2^0 + \cos\alpha_2 h_2^0\right)}{\sqrt{2}u'}$$

$$t_L t_R^c\ a_i^0:\quad \frac{im_t \sin\beta_2}{\sqrt{2}v'}\gamma_5 A_2^0;\quad b_L b_R^c\ a_i^0:\quad \frac{im_b \cos\beta_2}{\sqrt{2}u'}\gamma_5 A_2^0;$$

$$t_L b_R^c\ H_i^-:\quad m_b\frac{\left(\sin\gamma_2 H_2^- + \cos\gamma_2 h_2^-\right)}{\sqrt{2}u'};\quad b_L t_R^c\ H_i^+:\quad m_t\frac{\left(\cos\gamma_2 H_2^+ - \sin\gamma_2 h_2^+\right)}{\sqrt{2}v'}$$

### 6.1 Higgs-boson couplings to gauge bosons

We consider the interactions of Higgs scalars with gauge bosons in this section. The gauge invariant kinetic term is

$$L^{kin}_{Higgs} = (\Delta_\mu\varphi)^\dagger(\Delta^\mu\varphi) + (\Delta_\mu\varphi')^\dagger(\Delta_\mu\varphi')$$

The covariant derivative $\Delta_\mu$ in the kinetic term is defined model as [12]

$$\Delta_\mu = \partial_\mu - ig/2\, W_\mu^a T_a - ie A_\mu Q - \frac{ie}{s_W c_W}(T_{3L} - Q s_W^2) Z_\mu + \frac{ig}{c_W}\sqrt{(1-4s_W^2)}\,[T_{8L} + \frac{\sqrt{3}\, s_W^2}{(1-4s_W^2)} X] Z'_\mu$$

The trilinear and quartic vertices arise from the Lagrangian

- $L_{kin}^{Higgs} = L_{VVH} + L_{HHV} + L_{VVHH}$

$V_\mu$ represents the 15-plet gauge bosons of $SU(3)_L$. For neutral scalars $H_i = (h_i^0, H_i^0)$, $i = 1,2,3$

- $L_{VVH_1} = \frac{g^2}{2} v_\phi \left[ W_\mu^+ W^{-\nu} + Y_\mu^+ Y^{-\nu} + \frac{1}{2c_W^2}\left\{ Z_\mu Z^\nu + \frac{1}{\sqrt{3}} Z_\mu Z'^\nu + \frac{1}{3} Z'_\mu Z'^\nu \right\} \right] g_{VVH_1} H_1$

For $\tan\beta = u/v$, $v_\phi = v$,

$$g_{W^+W^-H_1} = \frac{\cos(\alpha_1 - \beta)}{\cos\beta} = g_{ZZH_1};\ g_{W^+W^-h_1} = -\frac{\sin(\alpha_1 - \beta)}{\cos\beta} = g_{ZZh_1};\ g_{Y^+Y^-H_1} = \cos\alpha_1;\ g_{Y^+Y^-h_1} = -\sin\alpha_1$$

$$g_{ZZ'H_1} = -\sqrt{1-4s_W^2}\cos\alpha_1 + \tan\beta\sin\alpha_1 \frac{(1+2s_W^2)}{\sqrt{1-4s_W^2}};\ g_{ZZ'h_1} = \sqrt{1-4s_W^2}\sin\alpha_1 + \tan\beta\cos\alpha_1 \frac{(1+2s_W^2)}{\sqrt{1-4s_W^2}};$$

- $L_{VVH_2} = \frac{g^2}{2} \left[ W_\mu^+ W^{-\nu} + Y_\mu^+ Y^{-\nu} + Y_\mu^{++} Y^{--\nu} + \frac{1}{2c_W^2}\left\{ Z_\mu Z^\nu + \frac{1}{\sqrt{3}} Z_\mu Z'^\nu + \frac{1}{3} Z'_\mu Z'^\nu \right\} \right] g_{VVH2} H_2$

For $u' = 1$, $v' = 1\,\text{GeV}$

$$g_{W^+W^-H_2} = \cos\alpha_2 + \sin\alpha_2 = g_{ZZH_2};\ g_{W^+W^-h_2} = -\sin\alpha_2 + \cos\alpha_2 = g_{ZZh_2};\ g_{Y^+Y^-H_1} = \cos\alpha_2;\ g_{Y^+Y^-h_2} = -\sin\alpha_2$$

$$g_{ZZ'H_2} = -\sqrt{1-4s_W^2}\cos\alpha_2 + \sin\alpha_2 \frac{(1+2s_W^2)}{\sqrt{1-4s_W^2}};\ g_{ZZ'h_1} = \sqrt{1-4s_W^2}\sin\alpha_2 + \cos\alpha_2 \frac{(1+2s_W^2)}{\sqrt{1-4s_W^2}};$$

- $L_{VVH_3} = \frac{g^2}{2}\left( Y_\mu^+ Y^{-\nu} + Y_\mu^{++} Y^{--\nu} + \frac{2s_W^2}{1-4s_W^2} Z'_\mu Z'^\nu \right) g_{VVH_3} H_3$

$$g_{Y^+Y^-H_3} = g_{Y^{++}Y^{--}H_3} = V\cos\alpha_3 + V'\sin\alpha_3 = g_{Z'Z'H} \; ; g_{Y^+Y^-h_3} = g_{Y^{++}Y^{--}h_3} = -V\sin\alpha_3 + V'\cos\alpha_3 = g_{Z'Z'h_3};$$

## 7. Result and Conclusions

We have proposed a supersymmetric version of a 3-3-1 model with exotic charged quarks and leptons. [11], which can be embedded in SU(4)$_C$⊗SU(4)$_W$ gauge symmetry[12]. The weak isospin group SU(2)$_L$ is a subgroup of a larger SU(3)$_L$⊗U(1)$_{Y1}$ symmetry group .The mass spectrum and couplings of the extended Higgs sector is studied.A possible set of parameters of the model are obtained which are consistent with other supersymmetric versions of 3-3-1 models[8].The lightest neutral scalar $h_1^0$, pseudoscalar $A_1$ and single-charged scalar $H_1^\pm$ are found to be in agreement with MSSM predictions with a trilinear soft term introduced in 3-3-1 models as $\varepsilon\rho\eta\chi$ [8]. This is distinct from a study of this model with only bilinear soft terms which allows very light Higgs masses[13] The exotic scalars and fermions are predicted to be at TeV energies .The spectrum for super particles will be presented elsewhere.

### 7.References

15. S.Sen and A.Dixit,[arXiv:hep-ph/0503078]

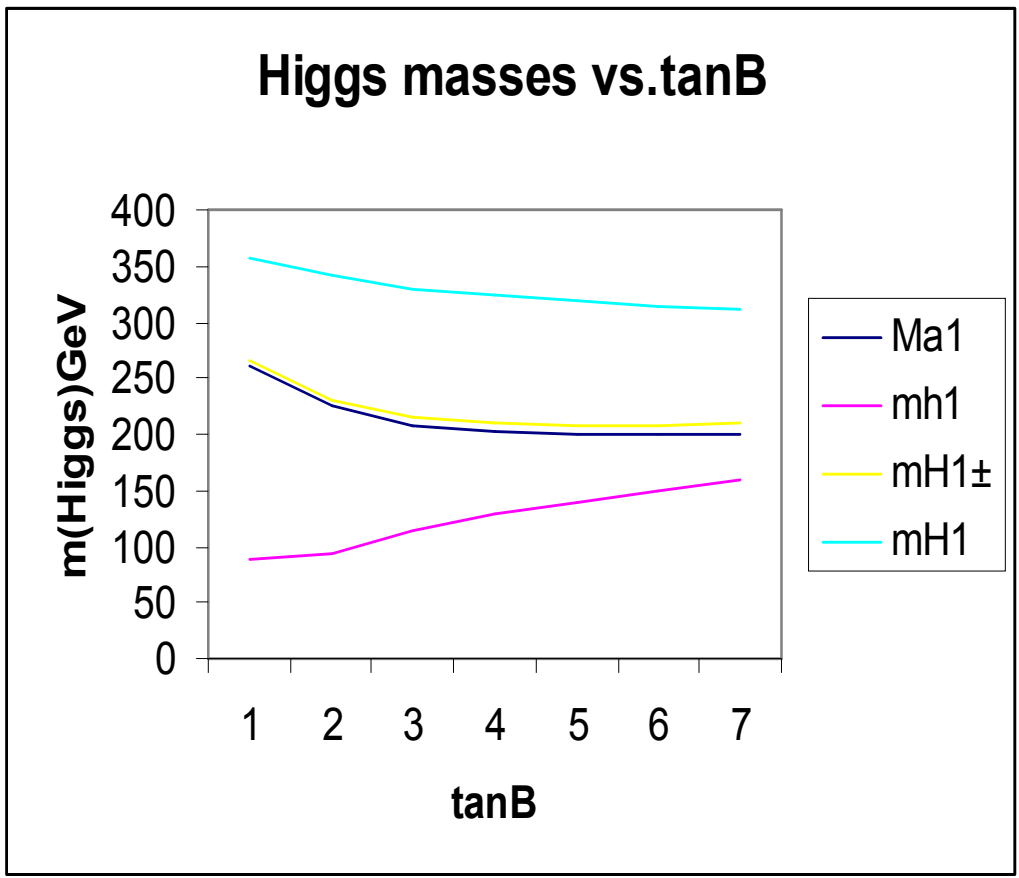

Fig.1:Higgs masses vs. $\tan\beta = u/v$ at V=1TeV

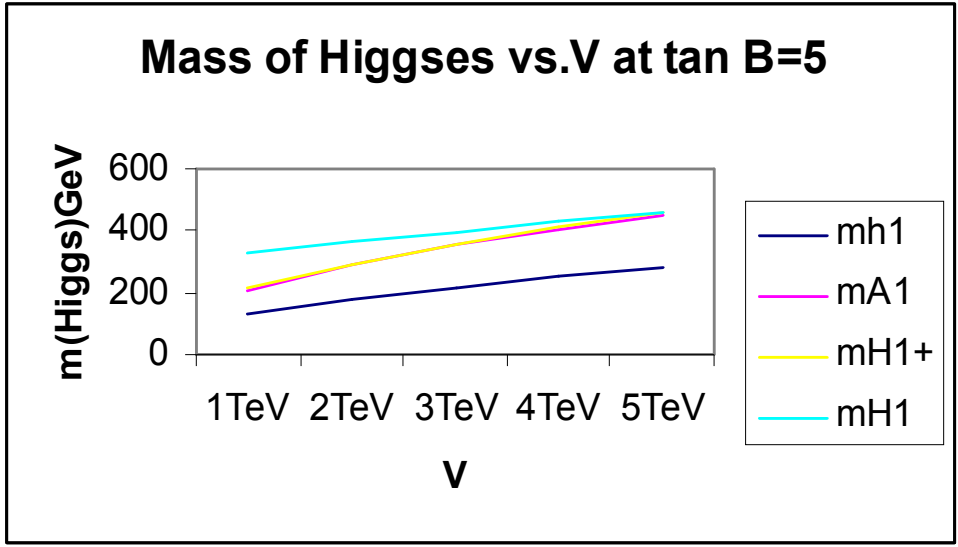

Fig.2: Higgs masses vs. V at tanβ = 5.

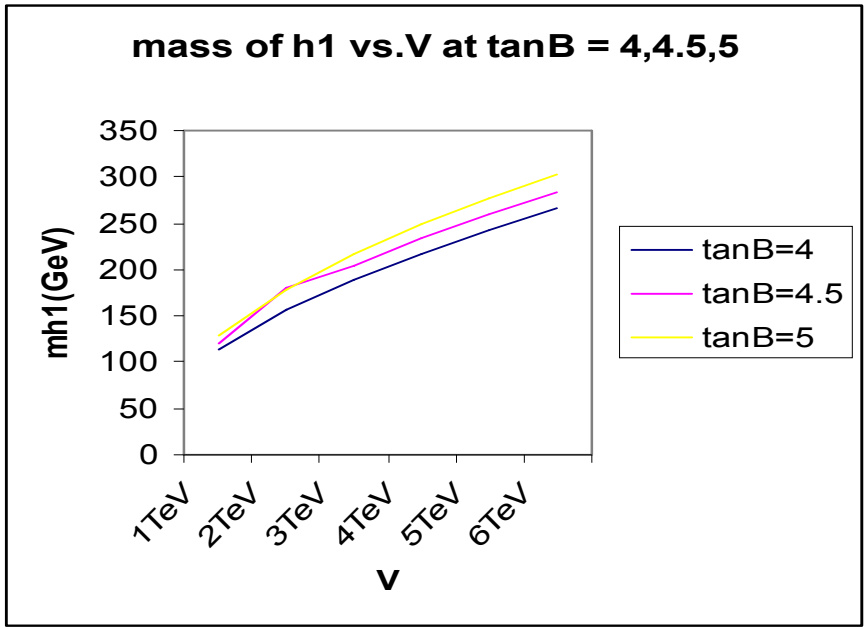

Fig.3: Mass of lightest Higgs scalar $h_1^0$ vs.V at $\tan\beta$ = 4,4.5 and 5

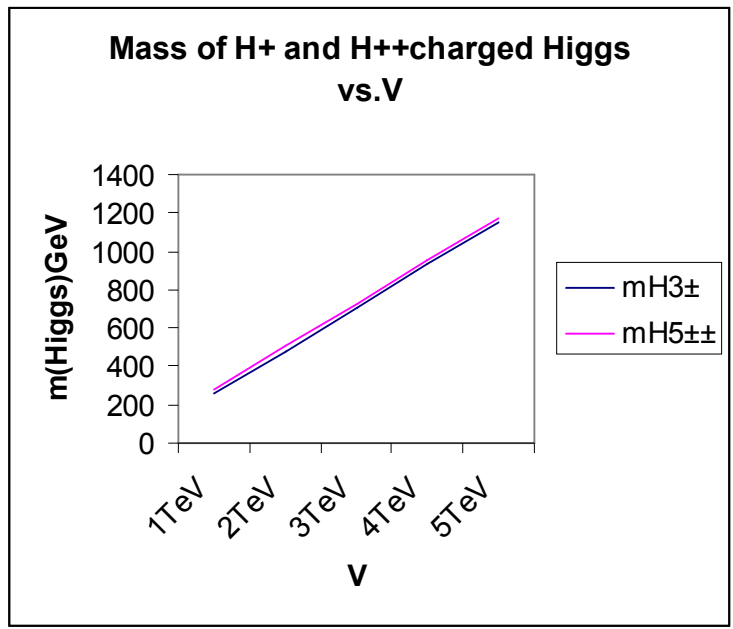

Fig.4: Higgs masses vs.V at tanβ=5 for $H_3^{\pm}$ and $H_5^{\pm\pm}$ charged scalars.